\documentclass[12pt]{article}
\usepackage{graphicx}

\begin{document}

\begin{titlepage}
\vskip 0.1cm

\centerline{\large \bf THE USE OF CONFIGURATION INTERACTION
METHOD}
 \centerline{\large \bf FOR DESCRIBING "FINE" - SPLITTING
EFFECTS}
\centerline{\large \bf  IN THE BOUND TWO-QUARK SYSTEM}

\vskip 0.7cm

\centerline{V. Lengyel, V. Rubish$^{\ast}$, A. Shpenik$^{\ddagger}$}

\vskip .2cm

\centerline{\sl Uzhgorod State University,}
\centerline{\sl Department of Theoretical Physics, Voloshin str. 32,}
\centerline{\sl 88000 Uzhgorod, Ukraine}

\vskip 0.7cm

\centerline{S.Chalupka}

\vskip .2cm

\centerline{\sl Safarik University,}
\centerline{\sl Department of Theoretical Physics and Geophysics, Moyzesova str. 16,}
\centerline{\sl 041 54 Kosice, Slovak Republic}

\vskip 0.7cm

\centerline{M.Salak}

\vskip .2cm

\centerline{\sl Presov University,}
\centerline{\sl Department of  Physics, 17 Novembra str.,}
\centerline{\sl 080 09 Presov, Slovak Republic}

\vskip 1.0cm

\begin{abstract}
The screened quasi-relativistic potential is used for describing
spin-orbit splitting in $^{3}P_{J}$ waves of quark-antiquark
system. Fermi-Breit equation is solved numerically in
configuration interaction approximation. This approximation takes
into account the mixing of wave functions up to fifth order and
corrects substantially perturbation calculations. The nature of
potentials Lorentz transformation property is elucidated. Very
good quantitative results for $b\overline{b}$ and $c\overline{c}$
quarkonia and quite acceptable qualitative characteristics as well
as for systems with unequal masses are obtained for
$u\overline{u}$.
\end{abstract}

\vskip .3cm

\vskip 4cm

\hrule

\vskip .3cm

\noindent
\vfill $ \begin{array}{ll} ^{\ast}\mbox{{\it e-mail
address:}} &
 \mbox{vrubish@univ.uzhgorod.ua}
\end{array}
$

$ \begin{array}{ll} ^{\ddagger}\mbox{{\it e-mail address:}} &
 \mbox{shpenik@iep.uzhgorod.ua}
\end{array}
$

\vfill
\end{titlepage}\eject
\baselineskip=14pt

Today it seems evident that quark potential model gives a rather
good description of spin-averaged mass spectrum of hadrons,
considered as a system of quarks (see e.g. \cite{1} and literature
cited there). Usually non-relativistic Cornell as well as
oscillator potential with Coulumb-like one-gluon exchange or other
power-law confinement terms is used. In this work we try to extend
this approach to incorporate the second order spin-terms in
two-quark Fermi-Breit equation for evaluating the spin-orbit
splitting. Instead of calculating as usual spin-terms in
first-order perturbation approximation the expansion of the total
wave function into a basic set of non-perturbative solutions up to
the fifth order of configurationally interacting states is carried
out.

The influence of a specific form of potential on a spin-effects (like
Cornel, oscillator, screened etc.) which generally give the same results for
spin-averaged data, the way of accounting of both long-range
non-perturbative and short-range (perturbative) parts in these calculations,
the Lorentz structure of different parts of potentials all these questions
remain to be unsolved yet.

Our main goal is to clarify some aspects of these questions in framework of
configuration interaction (CI) approach \cite{2}. This method does not need
the assumption that the coupling constant is to be small, the assumption
which is required by perturbation method. Since in quark potential case it
is not so the use of perturbation method looks very dubious.

Let us suggest that the static quark potential has vector and scalar
property of Lorentz transform:

\begin{equation}
V_{NR}\left( r\right) =V_{V}\left( r\right) +V_{S}\left( r\right) .
\label{f1}
\end{equation}

Index ''v'' means that potential is a 4-th component of operator
$\widehat{p}_{\mu }$, index ''s'' means that the potential is
scalar.

Following many authors we assume the admixture of vector-scalar screened
potential

\begin{equation}
V_{V}\left( r\right) =V_{OGE}+\varepsilon \cdot V_{Conf},
\label{f2}
\end{equation}

\begin{equation}
V_{S}\left( r\right) =(1-\varepsilon )\cdot V_{Conf}, \label{f3}
\end{equation}

where

$V_{OGE}-$ is the one gluon exchange potential,

$V_{Conf}-$ is the confinement part of the potential.

where $\varepsilon $ is mixing constant. Here the Lorentz nature
of the one-gluon and the confining potential is different, the
one-gluon potential being totally vector while the confining
potential is a vector-scalar mixture. This choice seems to be
reasonable since non-perturbative vertex corrections are important
at small $q^{2}$ i.e. large distances and naturally are coupled
only with the long-range term in potential. Very interesting
rewiew was done in the work of Nora Brambilla conserning the
choice of the interaction potential in \cite{3}.

In our case the choice of potential itself is dictated by the
consideration of most accurate description of averaged mass
spectrum, and here Chikovani-Jenkovszky-Paccanoni (CJP) type
potential \cite{4}.

Our approach is based on the model of a non-perturbative gluon propagator,
which was recently proposed by Chikovani, Jenkovszky and Paccanoni (CJP)
\cite{4}. In this model the propagator has the form

\begin{equation}
D\left( q^{2}\right) =\frac{c}{\left( q^{2}-\mu ^{2}\right)
}-\frac{1}{q^{2}-M^{2}}  \label{f4}
\end{equation}

and the corresponding potential assumes the form \cite{4}

\begin{equation}
V\left( r\right) =\frac{g^{2}}{6\pi \mu }\left( 1-\exp \left[ -\mu r\right]
\right) -\frac{16\pi }{25}\frac{\exp \left[ -Mr\right] }{r\cdot \ln \left(
b+1/\left( \Lambda r\right) ^{2}\right) }.  \label{f5}
\end{equation}

In fact for accelerating numerical calculations we use simple form
$\frac{\alpha _{s}}{r}$ for one-gluon-exchange type term with
QCD-like of asymptotic freedom

\begin{equation}
\alpha _{s}\left( r\right) =\frac{12\pi }{33-2N}\cdot \frac{1}{\ln \left(
1/\left( \widetilde{\Lambda }r\right) ^{2}\right) },  \label{f6}
\end{equation}

where $\widetilde{\Lambda }$ is taken to be equal to
$\widetilde{\Lambda }=0.14$ $GeV$. The initial value of $\alpha
_{s}$ was defined via calculating $q\overline{q}-$ masses.

At the finally, we present scalar and vector parts of the
potential (\ref{f2},\ref{f3}) in form

\begin{equation}
V_{OGE}\left( r\right) =-\frac{\alpha _{S}}{r},  \label{f7}
\end{equation}

\begin{equation}
V_{Conf}\left( r\right) =\frac{g^{2}}{6\pi }\cdot \frac{\left( 1-e^{-\mu
r}\right) }{\mu }  \label{f8}
\end{equation}

Recently screened potential of type (\ref{f7},\ref{f8}) was successfully
applied to a description of spin-averaged meson and baryon mass-spectra \cite
{5}. Recently, the study of the quark confining potential from QCD has been
well developed. For example, the consideration of the gluon condense may
give the confining potential a theoretical description. By considering the
loop-diagrams of fermion in the lattice gauge calculation, E. Laermann et al
\cite{6} showed that the quark confinement potential goes lower than the
linear confinement potential, when \textbf{r}, the distance between quarks,
becomes larger. It reflects the screening between valence quarks and quark
sea enhances, consequently, it suppresses the strength of color confinement
between valance quarks. According to quantum mechanics, this feature may
reduce the energies of highly excited baryons and mesons and help us to
understand the experimental data.

Timo Thonhauser \cite{7} approached the problem of the screened potential
from opposite site, namely of Few-Body approach to baryon structure.
Few-Body potential for baryons usually has confinement \cite{8} part plus
screened part. Very detailed description of nucleon-nucleon interaction is
obtained within this approach. Thonhauser applied this potential to meson
structure and obtained quite reasonable values of averaged meson masses.
Unfortunately he did not report the results concerning fine splitting. The
question of comparison of two approaches needs further study.

In addition, (see e. g. Gerasimov \cite{9}) the spin-orbit term has to be of
short range, as is indicated by quantum chromodinamics (QCD), the condition
which is evidently satisfied by CJP potential.

In what follows we shall use the screened potential \cite{4}, which proved
to be very good in describing the spin-averaged mass-spectrum of both bosons
and baryons as quark systems \cite{4},\cite{10}-\cite{13} and which secures
the necessary fall-of the spin-dependent forces.

Let us start with two-body Fermi-Breit equation. We shall use nuclear system
of units $\hbar =c=1$,$1GeV=\frac{5.068}{1Fm}.$ The Hamiltonian of the
system has the form:

\begin{equation}
\widehat{H}=\widehat{H}_{0}+\widehat{W},  \label{f9}
\end{equation}

where

\begin{equation}
\widehat{H}_{0}=-\frac{1}{2m}\triangle +\left( -\frac{\alpha
_{S}}{r}+\frac{g^{2}}{6\pi }\frac{\left( 1-e^{-\mu r}\right) }{\mu
}\right) ,  \label{f10}
\end{equation}

$m$ is the reduced mass,

\begin{equation}
\widehat{W}=\widehat{H}_{LS}+\widehat{H}_{ST},  \label{f11}
\end{equation}

where spin-dependent potentials are given by \cite{1}:

$\bullet$ Spin-orbit interaction

\begin{equation}
\begin{array}{c}
\widehat{H}_{LS}=\frac{1}{4m_{1}^{2}m_{2}^{2}}\frac{1}{r}\{\left[
((m_{1}+m_{2})^{2}+2m_{1}m_{2})\overrightarrow{L}\cdot
\overrightarrow{S_{+}}
+(m_{2}^{2}-m_{1}^{2})\overrightarrow{L}\cdot
\overrightarrow{S_{-}}\right] \frac{dV_{V}}{dr}- \\ -\left[ \left(
m_{1}^{2}+m_{2}^{2}\right) \overrightarrow{L}\cdot
\overrightarrow{S_{+}}+(m_{2}^{2}-m_{1}^{2})\overrightarrow{L}\cdot
\overrightarrow{S_{-}}\right] \frac{dV_{S}}{dr}\}, \\
\overrightarrow{L}=\overrightarrow{r}\times \overrightarrow{p,}
\qquad \overrightarrow{S_{+}}\equiv
\overrightarrow{S_{1}}+\overrightarrow{S_{2,}} \qquad
\overrightarrow{S_{-}}\equiv \overrightarrow{S_{1}}-
\overrightarrow{S_{2}}, \\ \overrightarrow{L}\cdot
\overrightarrow{S}=\frac{1}{2}\left[ j\left( j+1\right) -l\left(
l+1\right) -S\left( S+1\right) \right] .
\end{array}
\label{f12}
\end{equation}

Tensor terms

\begin{equation}
\begin{array}{c}
\widehat{H}_{T}=\frac{1}{12m_{1}m_{2}}\left[
\frac{1}{r}\frac{dV_{V}}{dr}- \frac{d^{2}V_{V}}{dr^{2}}\right]
\cdot S_{12}, \\ S_{12}=\frac{4}{\left( 2l+3\right) \left(
2l-1\right) }\left[ \overrightarrow{S}^{2}\cdot
\overrightarrow{L}^{2}-\frac{3}{2} \overrightarrow{L}\cdot
\overrightarrow{S}-3\left( \overrightarrow{L}\cdot
\overrightarrow{S}\right) ^{2}\right] .
\end{array}
\label{f13}
\end{equation}

All notations are considered to be familiar and coincide with those which
were used by Lucha and Schoberl in their well-known monograph \cite{1}.

Lucha and Shoberl \cite{1} has indicated that the term
$\frac{1}{r}\frac{dV_{V}}{dr}$ causes serious dimensional trouble
because of particle falling on the center. But in a more
consistent approach based on using Dirac equation we would have
instead of previous term the following one
$\frac{1}{E-V+mc^{2}}\frac{1}{r}\frac{dV_{V}}{dr}$, which for
$r\rightarrow 0$ will behave like $\frac{1}{r^{2}}$ and the
problem is removed. In matrix elements, we have calculated we
obtained that, there are no essential differences between these
two results. Previously we have studied these questions in
\cite{14}.

The ether possibility to solve this problem is the regularisation
of the potential \cite{1}. The same idea is based on the using of
$V_{reg}\propto \frac{1}{r+r_{0}}$ in the meaning of the
regularized potential; where $r_{0}- $ is free parameter, which
has been obtain from the comparison with the experimental data. So
in the case of $r=0$ the problem of particle fooling is gone.

Based on this reason we believe that our numerical results are not up set by
this problem.

However there are more ''exact'' QCD - motivated potentials, which are based
on two loop back diagrams \cite{15}. But by more detailed consideration it
appears, that these accounts were carried out only for one-gluon exchange
part. As races for systems considered by us, the very essential role is
played by confinement. Besides use of this potential results in increase of
number of parameters, with 1 in our case up to 3. Moreover analysis the
obtained results with two loopback potential show that of essential
improvement of concurrence theoretical and experimental data is not observed.

In addition to above, indicated terms presented in Fermi-Breit
Hamiltonian are the $\overrightarrow{S_{1}}\cdot
\overrightarrow{S_{2}}$ (spin-spin) and relativistic correction
terms (of $p^{4}$-order). Some authors (like \cite{1}) indicated
that these terms are important for calculating mass spectra, other
authors (like \cite{16}) argue that these results are purely
constant. In our case we believe that for LS-mass differences they
will not play an essential role.

Now, we consider Fermi-Breit equation

\begin{equation}
\left( -\frac{1}{2m}\triangle +\left( -\frac{\alpha
_{S}}{r}+\frac{g^{2}}{6\pi }\frac{\left( 1-e^{-\mu r}\right) }{\mu
}\right) +\widehat{W}\right) \Psi \left( \overrightarrow{r}\right)
=\widehat{E}\Psi \left( \overrightarrow{r}\right).  \label{f14}
\end{equation}

There are certain difficulties in solving equation (\ref{f14}) in the case
of spin-orbit-coupling. Using the potential (\ref{f11}) lead us to the
appearance of term of order $1/r^{3}$. Usually authors use one of two
possibilities:

a) use the perturbation method;

b) use the numerical computation.

Both of these possibilities are inferior: first one is unacceptable because
in some cases the ''fine'' splitting turns out to be not so ''fine'' at all,
being up to 50\% contribution to the final mass, the second one needs
introducing the cut-off parameter which is highly undesirable.

Here we suggest to use CI approach which was previously very successfully
applied in atomic physics \cite{2}. The essence of this approximation is
that the total wave function $\Psi \left( \overrightarrow{r}\right) $ is
expanded in set of eigen functions $\varphi _{n}$of the unperturbated
Hamiltonian $\widehat{H}_{0}$, that is

\begin{equation}
\Psi \left( \overrightarrow{r}\right) =\sum a_{n}\varphi
_{n}\left( \overrightarrow{r}\right).  \label{f15}
\end{equation}

After substituting (\ref{f15}) into (\ref{f14}) and using eigen
value $E_{n}^{0}$ we obtain the system of linear equations for
$a_{n}$ which have to be truncated for reasonably large $n$.

\begin{equation}
\begin{array}{c}
a_{1}\left( E-E_{1}^{0}-W_{11}\right) -a_{2}W_{12}-a_{3}W_{13}-\ldots
-a_{n}W_{1n}=0 \\
-a_{1}W_{21}+a_{1}\left( E-E_{2}^{0}-W_{22}\right) -a_{3}W_{23}-\ldots
-a_{n}W_{2n}=0 \\
\cdots \cdots \cdots \cdots \cdots \cdots \cdots \cdots \cdots \cdots \cdots
\cdots \cdots \cdots \cdots \cdots \cdots \cdots \cdots \\
-a_{1}W_{n1}-a_{2}W_{n2}-a_{3}W_{n3}-\ldots +a_{n}\left(
E-E_{n}^{0}-W_{nn}\right) =0
\end{array}
,  \label{f16}
\end{equation}

where

\begin{equation}
W_{ij}=\left\langle \varphi _{i}\right| \widehat{W}\left| \varphi
_{j}\right\rangle .  \label{f17}
\end{equation}

Both the basic functions $\varphi _{i}$ and matrix elements $W_{ij}$ are
calculated numerically. Nontrivial solution there will be only if the
determinant of this system vanishes can be solved by diagonalizing of the
matrix for $E$. The system (\ref{f16}) is called CI. This procedure goes far
outside of the perturbation method.

The CI approximation turned out to be extremely successful in
atomic physics. In calculating atomic structure it allowed to
increase the precision of calculating energy levels by one order.
In the scattering processes it allowed to reveal fine resonance
structure in scattering cross-sections due to formation of
auto-ionizing states. So we expect that its applications will be
even more important in strong interaction, where the perturbation
method is evidently not correct. The technique of application of
CIA is quite complicated, since it needs to handle the matrices of
large dimensions. In current work we used the code elaborated by
O. Zatsarinny \cite{17}.

In this work we have applied the above described method for
calculating $P-$ wave ''fine''-splitting of $b\overline{b}-$,
$c\overline{c}-$ and $u \overline{u}-$ systems to $^{3}P_{0}$,
$^{3}P_{1}$ and $^{3}P_{2}$ levels. As well as for quarkonium
constituted out of different masses like $s \overline{u}$, etc. In
our case the corresponding operators $\widehat{H}_{LS} $,
$\widehat{H}_{T}$ will be has the form

\begin{equation}
\widehat{W}_{LS}=\frac{1}{2m^{2}}\frac{1}{r}\left[ 3\frac{\alpha _{S}}{r^{2}}%
+\left( 4\varepsilon -1\right) \cdot \frac{g^{2}}{6\pi }e^{-\mu r}\right]
\overrightarrow{L}\cdot \overrightarrow{S},  \label{f18}
\end{equation}

\begin{equation}
\widehat{W}_{T}=\frac{1}{12m^{2}}\left[ 3\frac{\alpha _{S}}{r^{3}}+\left(
\frac{1}{r}-\mu \right) \varepsilon \cdot \frac{g^{2}}{6\pi }e^{-\mu
r}\right] S_{12}.  \label{f19}
\end{equation}

It is important that all parameters except $\varepsilon $ are
taken from \cite{4},\cite{13}, where excellent description of
bottomonium and charmonium spectra were obtained. Moreover as it
was shown in \cite{13} the same parameters gives good $\rho -$
meson trajectories. Actually the values $\frac{g^{2}}{6\pi }=0.3$
$GeV^{2}$, $\mu =0.054$ $GeV$ were taken and $\alpha _{S}$ was
taken in accordance to QCD. The only adjustable parameter was
$\varepsilon $. The experimental values are taken from \cite{18}.
As mentioned above all calculations were carried out numerically.
Special code was constructed for this purpose. The calculations
were extended to fifth order in (\ref{f15}) ( see tables, rows
1-8), i.e. until the differences between the results did not go
below several $MeV$ level.

\medskip

Table 1. $b\overline{b}$-system, $\alpha _{s}=0.3$, $\varepsilon
=0.5$, $m_{b}=5.05$ $GeV.$

\begin{tabular}{|c|c|c|c|c|}
\hline
\textbf{State} & $\mathbf{\Delta M}_{TH}$ & \textbf{order 1} & \textbf{order
2} & $\mathbf{\Delta M}_{EXP}[18]$ \\
&  & $\mathbf{MeV}$ & $\mathbf{MeV}$ & $\mathbf{MeV}$ \\ \hline
$\chi _{b2}\left( 0^{+}\left( 2^{++}\right) \right) -\chi _{b1}\left(
0^{+}\left( 1^{++}\right) \right) $ & $1^{3}P_{2}-1^{3}P_{1}$ & $21.32$ & $%
21.23$ & $21.3\pm 1.3$ \\ \hline
$\chi _{b1}\left( 0^{+}\left( 1^{++}\right) \right) -\chi _{b0}\left(
0^{+}\left( 0^{++}\right) \right) $ & $1^{3}P_{1}-1^{3}P_{0}$ & $24.25$ & $%
25.51$ & $32.1\pm 2$ \\ \hline
$\chi _{b2}\left( 0^{+}\left( 2^{++}\right) \right) -\chi _{b0}\left(
0^{+}\left( 0^{++}\right) \right) $ & $1^{3}P_{2}-1^{3}P_{0}$ & $45.57$ & $%
46.74$ & $53.4\pm 1.9$ \\ \hline
$R$ &  & $0.88$ & $0.83$ & $0.66$ \\ \hline
$\chi _{b2}\left( 0^{+}\left( 2^{++}\right) \right) -\chi _{b1}\left(
0^{+}\left( 1^{++}\right) \right) $ & $2^{3}P_{2}-2^{3}P_{1}$ & $-$ & $17.72$
& $13.3\pm 0.9$ \\ \hline
$\chi _{b1}\left( 0^{+}\left( 1^{++}\right) \right) -\chi _{b0}\left(
0^{+}\left( 0^{++}\right) \right) $ & $2^{3}P_{1}-2^{3}P_{0}$ & $-$ & $19.09$
& $23.1\pm 1.1$ \\ \hline
$\chi _{b2}\left( 0^{+}\left( 2^{++}\right) \right) -\chi _{b0}\left(
0^{+}\left( 0^{++}\right) \right) $ & $2^{3}P_{2}-2^{3}P_{0}$ & $-$ & $36.81$
& $36.4\pm 1.0$ \\ \hline
$R$ &  & $-$ & $0.93$ & $0.57$ \\ \hline
\end{tabular}

\begin{tabular}{|c|c|c|c|c|}
\hline
\textbf{State} & $\mathbf{\Delta M}_{TH}$ & \textbf{order 3} & \textbf{order
4} & $\mathbf{\Delta M}_{EXP}[18]$ \\
&  & $\mathbf{MeV}$ & $\mathbf{MeV}$ & $\mathbf{MeV}$ \\ \hline
$\chi _{b2}\left( 0^{+}\left( 2^{++}\right) \right) -\chi _{b1}\left(
0^{+}\left( 1^{++}\right) \right) $ & $1^{3}P_{2}-1^{3}P_{1}$ & $21.20$ & $%
21.19$ & $21.3\pm 1.3$ \\ \hline
$\chi _{b1}\left( 0^{+}\left( 1^{++}\right) \right) -\chi _{b0}\left(
0^{+}\left( 0^{++}\right) \right) $ & $1^{3}P_{1}-1^{3}P_{0}$ & $26.08$ & $%
26.44$ & $32.1\pm 2$ \\ \hline
$\chi _{b2}\left( 0^{+}\left( 2^{++}\right) \right) -\chi _{b0}\left(
0^{+}\left( 0^{++}\right) \right) $ & $1^{3}P_{2}-1^{3}P_{0}$ & $47.29$ & $%
47.63$ & $53.4\pm 1.9$ \\ \hline
$R$ &  & $0.81$ & $0.8$ & $0.66$ \\ \hline
$\chi _{b2}\left( 0^{+}\left( 2^{++}\right) \right) -\chi _{b1}\left(
0^{+}\left( 1^{++}\right) \right) $ & $2^{3}P_{2}-2^{3}P_{1}$ & $17.61$ & $%
17.58$ & $13.3\pm 0.9$ \\ \hline
$\chi _{b1}\left( 0^{+}\left( 1^{++}\right) \right) -\chi _{b0}\left(
0^{+}\left( 0^{++}\right) \right) $ & $2^{3}P_{1}-2^{3}P_{0}$ & $20.37$ & $21
$ & $23.1\pm 1.1$ \\ \hline
$\chi _{b2}\left( 0^{+}\left( 2^{++}\right) \right) -\chi _{b0}\left(
0^{+}\left( 0^{++}\right) \right) $ & $2^{3}P_{2}-2^{3}P_{0}$ & $37.98$ & $%
38.58$ & $36.4\pm 1.0$ \\ \hline
$R$ &  & $0.86$ & $0.84$ & $0.57$ \\ \hline
\end{tabular}

\newpage

Table 2. $b\overline{b}$-system, $\alpha _{s}=0.3$, $\varepsilon =0.45$, $%
m_{b}=5.05$ $GeV.$

\begin{tabular}{|c|c|c|c|c|}
\hline
\textbf{State} & $\mathbf{\Delta M}_{TH}$ & \textbf{order 1} & \textbf{order
2} & $\mathbf{\Delta M}_{EXP}[18]$ \\
&  & $\mathbf{MeV}$ & $\mathbf{MeV}$ & $\mathbf{MeV}$ \\ \hline
$\chi _{b2}\left( 0^{+}\left( 2^{++}\right) \right) -\chi _{b1}\left(
0^{+}\left( 1^{++}\right) \right) $ & $1^{3}P_{2}-1^{3}P_{1}$ & $20.00$ & $%
19.92$ & $21.3\pm 1.3$ \\ \hline
$\chi _{b1}\left( 0^{+}\left( 1^{++}\right) \right) -\chi _{b0}\left(
0^{+}\left( 0^{++}\right) \right) $ & $1^{3}P_{1}-1^{3}P_{0}$ & $23.37$ & $%
24.57$ & $32.1\pm 2$ \\ \hline
$\chi _{b2}\left( 0^{+}\left( 2^{++}\right) \right) -\chi _{b0}\left(
0^{+}\left( 0^{++}\right) \right) $ & $1^{3}P_{2}-1^{3}P_{0}$ & $43.37$ & $%
44.49$ & $53.4\pm 1.9$ \\ \hline
$R$ &  & $0.85$ & $0.81$ & $0.66$ \\ \hline
$\chi _{b2}\left( 0^{+}\left( 2^{++}\right) \right) -\chi _{b1}\left(
0^{+}\left( 1^{++}\right) \right) $ & $2^{3}P_{2}-2^{3}P_{1}$ & $-$ & $16.68$
& $13.3\pm 0.9$ \\ \hline
$\chi _{b1}\left( 0^{+}\left( 1^{++}\right) \right) -\chi _{b0}\left(
0^{+}\left( 0^{++}\right) \right) $ & $2^{3}P_{1}-2^{3}P_{0}$ & $-$ & $18.49$
& $23.1\pm 1.1$ \\ \hline
$\chi _{b2}\left( 0^{+}\left( 2^{++}\right) \right) -\chi _{b0}\left(
0^{+}\left( 0^{++}\right) \right) $ & $2^{3}P_{2}-2^{3}P_{0}$ & $-$ & $35.25$
& $36.4\pm 1.0$ \\ \hline
$R$ &  & $-$ & $0.9$ & $0.57$ \\ \hline
\end{tabular}

\begin{tabular}{|c|c|c|c|c|}
\hline
\textbf{State} & $\mathbf{\Delta M}_{TH}$ & \textbf{order 3} & \textbf{order
4} & $\mathbf{\Delta M}_{EXP}[18]$ \\
&  & $\mathbf{MeV}$ & $\mathbf{MeV}$ & $\mathbf{MeV}$ \\ \hline
$\chi _{b2}\left( 0^{+}\left( 2^{++}\right) \right) -\chi _{b1}\left(
0^{+}\left( 1^{++}\right) \right) $ & $1^{3}P_{2}-1^{3}P_{1}$ & $19.89$ & $%
19.87$ & $21.3\pm 1.3$ \\ \hline
$\chi _{b1}\left( 0^{+}\left( 1^{++}\right) \right) -\chi _{b0}\left(
0^{+}\left( 0^{++}\right) \right) $ & $1^{3}P_{1}-1^{3}P_{0}$ & $25.13$ & $%
25.48$ & $32.1\pm 2$ \\ \hline
$\chi _{b2}\left( 0^{+}\left( 2^{++}\right) \right) -\chi _{b0}\left(
0^{+}\left( 0^{++}\right) \right) $ & $1^{3}P_{2}-1^{3}P_{0}$ & $45.02$ & $%
45.36$ & $53.4\pm 1.9$ \\ \hline
$R$ &  & $0.79$ & $0.78$ & $0.66$ \\ \hline
$\chi _{b2}\left( 0^{+}\left( 2^{++}\right) \right) -\chi _{b1}\left(
0^{+}\left( 1^{++}\right) \right) $ & $2^{3}P_{2}-2^{3}P_{1}$ & $16.76$ & $%
16.65$ & $13.3\pm 0.9$ \\ \hline
$\chi _{b1}\left( 0^{+}\left( 1^{++}\right) \right) -\chi _{b0}\left(
0^{+}\left( 0^{++}\right) \right) $ & $2^{3}P_{1}-2^{3}P_{0}$ & $19.73$ & $%
20.34$ & $23.1\pm 1.1$ \\ \hline
$\chi _{b2}\left( 0^{+}\left( 2^{++}\right) \right) -\chi _{b0}\left(
0^{+}\left( 0^{++}\right) \right) $ & $2^{3}P_{2}-2^{3}P_{0}$ & $36.38$ & $%
36.95$ & $36.4\pm 1.0$ \\ \hline
$R$ &  & $0.85$ & $0.82$ & $0.57$ \\ \hline
\end{tabular}

\bigskip

\bigskip

\bigskip

Table 3. $c\overline{c}$-system, $\alpha _{s}=0.386$, $\varepsilon =0.3$, $%
m_{c}=1.675$ $GeV.$

\begin{tabular}{|c|c|c|c|c|c|}
\hline
\textbf{State} & $\mathbf{\Delta M}_{TH}$ & \textbf{order 1} & \textbf{order
2} & \textbf{order 3} & \textbf{order 4} \\
&  & $\mathbf{MeV}$ & $\mathbf{MeV}$ & $\mathbf{MeV}$ & $\mathbf{MeV}$ \\
\hline
$\chi _{c2}\left( 0^{+}\left( 2^{++}\right) \right) -\chi _{c1}\left(
0^{+}\left( 1^{++}\right) \right) $ & $1^{3}P_{2}-1^{3}P_{1}$ & $52.20$ & $%
51.66$ & $51.49$ & $51.41$ \\ \hline
$\chi _{c1}\left( 0^{+}\left( 1^{++}\right) \right) -\chi _{c0}\left(
0^{+}\left( 0^{++}\right) \right) $ & $1^{3}P_{1}-1^{3}P_{0}$ & $69.31$ & $%
77.27$ & $81.55$ & $84.51$ \\ \hline
$\chi _{c2}\left( 0^{+}\left( 2^{++}\right) \right) -\chi _{c2}\left(
0^{+}\left( 0^{++}\right) \right) $ & $1^{3}P_{2}-1^{3}P_{0}$ & $121.51$ & $%
128.93$ & $133.04$ & $135.93$ \\ \hline
$R$ &  & $0.75$ & $0.66$ & $0.63$ & $0.61$ \\ \hline
\end{tabular}

\begin{tabular}{|c|c|c|c|c|}
\hline
\textbf{State} & $\mathbf{\Delta M}_{TH}$ & \textbf{order 5} & \textbf{order
6} & $\mathbf{\Delta M}_{EXP}[18]$ \\
&  & $\mathbf{MeV}$ & $\mathbf{MeV}$ & $\mathbf{MeV}$ \\ \hline
$\chi _{c2}\left( 0^{+}\left( 2^{++}\right) \right) -\chi _{c1}\left(
0^{+}\left( 1^{++}\right) \right) $ & $1^{3}P_{2}-1^{3}P_{1}$ & $51.37$ & $%
51.33$ & $45.64\pm 0.25$ \\ \hline
$\chi _{c1}\left( 0^{+}\left( 1^{++}\right) \right) -\chi _{c0}\left(
0^{+}\left( 0^{++}\right) \right) $ & $1^{3}P_{1}-1^{3}P_{0}$ & $86.72$ & $%
88.48$ & $95.43\pm 1.12$ \\ \hline
$\chi _{c2}\left( 0^{+}\left( 2^{++}\right) \right) -\chi _{c2}\left(
0^{+}\left( 0^{++}\right) \right) $ & $1^{3}P_{2}-1^{3}P_{0}$ & $138.09$ & $%
139.82$ & $141.07\pm 1.13$ \\ \hline
$R$ &  & $0.59$ & $0.58$ & $0.48$ \\ \hline
\end{tabular}

\newpage

Table 4. $c\overline{c}$-system, $\alpha _{s}=0.386$, $\varepsilon =0.4$, $%
m_{c}=1.675$ $GeV.$

\begin{tabular}{|c|c|c|c|c|}
\hline
\textbf{State} & $\mathbf{\Delta M}_{TH}$ & \textbf{order 1} & \textbf{order
2} & $\mathbf{\Delta M}_{EXP}[18]$ \\
&  & $\mathbf{MeV}$ & $\mathbf{MeV}$ & $\mathbf{MeV}$ \\ \hline
$\chi _{c2}\left( 0^{+}\left( 2^{++}\right) \right) -\chi _{c1}\left(
0^{+}\left( 1^{++}\right) \right) $ & $1^{3}P_{2}-1^{3}P_{1}$ & $66.81$ & $%
66.17$ & $45.64\pm 0.25$ \\ \hline
$\chi _{c1}\left( 0^{+}\left( 1^{++}\right) \right) -\chi _{c0}\left(
0^{+}\left( 0^{++}\right) \right) $ & $1^{3}P_{1}-1^{3}P_{0}$ & $79.36$ & $%
88.57$ & $95.43\pm 1.12$ \\ \hline
$\chi _{c2}\left( 0^{+}\left( 2^{++}\right) \right) -\chi _{c2}\left(
0^{+}\left( 0^{++}\right) \right) $ & $1^{3}P_{2}-1^{3}P_{0}$ & $146.17$ & $%
154.74$ & $141.07\pm 1.13$ \\ \hline
$R$ &  & $0.84$ & $0.75$ & $0.48$ \\ \hline
\end{tabular}

\begin{tabular}{|c|c|c|c|c|}
\hline
\textbf{State} & $\mathbf{\Delta M}_{TH}$ & \textbf{order 3} & \textbf{order
4} & $\mathbf{\Delta M}_{EXP}[18]$ \\
&  & $\mathbf{MeV}$ & $\mathbf{MeV}$ & $\mathbf{MeV}$ \\ \hline
$\chi _{c2}\left( 0^{+}\left( 2^{++}\right) \right) -\chi _{c1}\left(
0^{+}\left( 1^{++}\right) \right) $ & $1^{3}P_{2}-1^{3}P_{1}$ & $65.99$ & $%
65.92$ & $45.64\pm 0.25$ \\ \hline
$\chi _{c1}\left( 0^{+}\left( 1^{++}\right) \right) -\chi _{c0}\left(
0^{+}\left( 0^{++}\right) \right) $ & $1^{3}P_{1}-1^{3}P_{0}$ & $93.45$ & $%
96.75$ & $95.43\pm 1.12$ \\ \hline
$\chi _{c2}\left( 0^{+}\left( 2^{++}\right) \right) -\chi _{c2}\left(
0^{+}\left( 0^{++}\right) \right) $ & $1^{3}P_{2}-1^{3}P_{0}$ & $159.45$ & $%
162.67$ & $141.07\pm 1.13$ \\ \hline
$R$ &  & $0.70$ & $0.68$ & $0.48$ \\ \hline
\end{tabular}

\bigskip

Table 5. $u\overline{u}$-system, $\alpha _{s}=0.52$, $\varepsilon =0.14$, $%
m_{u}=0.33$ $GeV.$\

\begin{tabular}{|c|c|c|c|c|}
\hline
\textbf{State} & $\mathbf{\Delta M}_{TH}$ & \textbf{order 1} & \textbf{order
2} & $\mathbf{\Delta M}_{EXP}[18]$ \\
&  & $\mathbf{MeV}$ & $\mathbf{MeV}$ & $\mathbf{MeV}$ \\ \hline
$a_{2}\left( 1^{-}\left( 2^{++}\right) \right) -a_{1}\left( 1^{-}\left(
1^{++}\right) \right) $ & $1^{3}P_{2}-1^{3}P_{1}$ & $15.34$ & $10.26$ & $%
88.2\pm 40.7$ \\ \hline
$a_{1}\left( 1^{-}\left( 2^{++}\right) \right) -a_{0}\left( 1^{-}\left(
0^{++}\right) \right) $ & $1^{3}P_{1}-1^{3}P_{0}$ & $241.19$ & $316.91$ & $%
246.5\pm 40.9$ \\ \hline
$a_{2}\left( 1^{-}\left( 2^{++}\right) \right) -a_{0}\left( 1^{-}\left(
0^{++}\right) \right) $ & $1^{3}P_{2}-1^{3}P_{0}$ & $256.53$ & $327.17$ & $%
334.6\pm 1.6$ \\ \hline
$R$ &  & $0.063$ & $0.032$ & $0.357$ \\ \hline
\end{tabular}

\begin{tabular}{|c|c|c|c|}
\hline
\textbf{State} & $\mathbf{\Delta M}_{TH}$ & \textbf{order 3} & $\mathbf{%
\Delta M}_{EXP}[18]$ \\
&  & $\mathbf{MeV}$ & $\mathbf{MeV}$ \\ \hline
$a_{2}\left( 1^{-}\left( 2^{++}\right) \right) -a_{1}\left( 1^{-}\left(
1^{++}\right) \right) $ & $1^{3}P_{2}-1^{3}P_{1}$ & $8.57$ & $88.2\pm 40.7$
\\ \hline
$a_{1}\left( 1^{-}\left( 2^{++}\right) \right) -a_{0}\left( 1^{-}\left(
0^{++}\right) \right) $ & $1^{3}P_{1}-1^{3}P_{0}$ & $374.09$ & $246.5\pm 40.9
$ \\ \hline
$a_{2}\left( 1^{-}\left( 2^{++}\right) \right) -a_{0}\left( 1^{-}\left(
0^{++}\right) \right) $ & $1^{3}P_{2}-1^{3}P_{0}$ & $382.66$ & $334.6\pm 1.6$
\\ \hline
$R$ &  & $0.023$ & $0.357$ \\ \hline
\end{tabular}

\bigskip

Table 6. $u\overline{u}$-system, $\alpha _{s}=0.52$, $\varepsilon =0.145$, $%
m_{u}=0.33$ $GeV.$

\begin{tabular}{|c|c|c|c|c|}
\hline
\textbf{State} & $\mathbf{\Delta M}_{TH}$ & \textbf{order 1} & \textbf{order
2} & $\mathbf{\Delta M}_{EXP}[18]$ \\
&  & $\mathbf{MeV}$ & $\mathbf{MeV}$ & $\mathbf{MeV}$ \\ \hline
$a_{2}\left( 1^{-}\left( 2^{++}\right) \right) -a_{1}\left( 1^{-}\left(
1^{++}\right) \right) $ & $1^{3}P_{2}-1^{3}P_{1}$ & $24.41$ & $19.15$ & $%
88.2\pm 40.7$ \\ \hline
$a_{1}\left( 1^{-}\left( 2^{++}\right) \right) -a_{0}\left( 1^{-}\left(
0^{++}\right) \right) $ & $1^{3}P_{1}-1^{3}P_{0}$ & $247.52$ & $324.96$ & $%
246.5\pm 40.9$ \\ \hline
$a_{2}\left( 1^{-}\left( 2^{++}\right) \right) -a_{0}\left( 1^{-}\left(
0^{++}\right) \right) $ & $1^{3}P_{2}-1^{3}P_{0}$ & $271.92$ & $344.10$ & $%
334.6\pm 1.6$ \\ \hline
$R$ &  & $0.098$ & $0.059$ & $0.357$ \\ \hline
\end{tabular}

\begin{tabular}{|c|c|c|c|}
\hline
\textbf{State} & $\mathbf{\Delta M}_{TH}$ & \textbf{order 3} & $\mathbf{%
\Delta M}_{EXP}[18]$ \\
&  & $\mathbf{MeV}$ & $\mathbf{MeV}$ \\ \hline
$a_{2}\left( 1^{-}\left( 2^{++}\right) \right) -a_{1}\left( 1^{-}\left(
1^{++}\right) \right) $ & $1^{3}P_{2}-1^{3}P_{1}$ & $17.45$ & $88.2\pm 40.7$
\\ \hline
$a_{1}\left( 1^{-}\left( 2^{++}\right) \right) -a_{0}\left( 1^{-}\left(
0^{++}\right) \right) $ & $1^{3}P_{1}-1^{3}P_{0}$ & $383.24$ & $246.5\pm 40.9
$ \\ \hline
$a_{2}\left( 1^{-}\left( 2^{++}\right) \right) -a_{0}\left( 1^{-}\left(
0^{++}\right) \right) $ & $1^{3}P_{2}-1^{3}P_{0}$ & $400.69$ & $334.6\pm 1.6$
\\ \hline
$R$ &  & $0.046$ & $0.357$ \\ \hline
\end{tabular}

\newpage

Table 7. $s\overline{u}$-system, $\alpha _{s}=0.421$, $\varepsilon =0.1875$,
$m_{s}=0.5$ $GeV,m_{u}=0.33$ $GeV.$

\begin{tabular}{|c|c|c|c|c|c|c|}
\hline
\textbf{State} & $\mathbf{\Delta M}_{TH}$ & \textbf{order 1} & \textbf{order
2} & \textbf{order 3} & \textbf{order 4} & $\mathbf{\Delta M}_{EXP}[18]$ \\
&  & $\mathbf{MeV}$ & $\mathbf{MeV}$ & $\mathbf{MeV}$ & $\mathbf{MeV}$ & $%
\mathbf{MeV}$ \\ \hline
$?$ & $1^{3}P_{2}-1^{3}P_{1}$ & $43.16$ & $40.23$ & $39.29$ & $38.87$ & $-$
\\ \hline
$?$ & $1^{3}P_{1}-1^{3}P_{0}$ & $187.78$ & $230.62$ & $259.86$ & $282.63$ & $%
-$ \\ \hline
$?$ & $1^{3}P_{2}-1^{3}P_{0}$ & $230.94$ & $270.85$ & $299.15$ & $321.50$ & $%
-$ \\ \hline
$R$ &  & $0.23$ & $0.17$ & $0.15$ & $0.14$ & $-$ \\ \hline
\end{tabular}

\bigskip

Table 8. $s\overline{u}$-system, $\alpha _{s}=0.421$, $\varepsilon =0.2$, $%
m_{s}=0.5$ $GeV,m_{u}=0.33$ $GeV.$

\begin{tabular}{|c|c|c|c|c|c|}
\hline
\textbf{State} & $\mathbf{\Delta M}_{TH}$ & \textbf{order 1} & \textbf{order
2} & \textbf{order 3} & $\mathbf{\Delta M}_{EXP}[18]$ \\
&  & $\mathbf{MeV}$ & $\mathbf{MeV}$ & $\mathbf{MeV}$ & $\mathbf{MeV}$ \\
\hline
$?$ & $1^{3}P_{2}-1^{3}P_{1}$ & $60.13$ & $56.95$ & $56.00$ & $-$ \\ \hline
$?$ & $1^{3}P_{1}-1^{3}P_{0}$ & $199.43$ & $244.92$ & $331.74$ & $-$ \\
\hline
$?$ & $1^{3}P_{2}-1^{3}P_{0}$ & $259.56$ & $301.84$ & $387.75$ & $-$ \\
\hline
$R$ &  & $0.30$ & $0.23$ & $0.17$ & $-$ \\ \hline
\end{tabular}

\newpage

{\large \bf Let us make the following conclusions:}

\bigskip

1. The results for heavy quarkonium are quite good for values
$\varepsilon =0.3-0.45$. Which is coincide with results obtained
by Lai-Him Chan \cite{19} For light quarkonium the results are
worse, which means that relativistic effects have to be taken into
account more carefully. We believe, that the difference between
$\varepsilon $ in case of $b\overline{b}$ and $u\overline{u}$
system exactly reflects this fact.

2. The value of indicates that confinement has prevailingly scalar
character. This conclusion does not contradict other authors
\cite{20}.

3. As it follows from (\ref{f18}) at $\varepsilon =0.25$ $\left(
4\varepsilon -1\right) =0$ the contribution of confinement vanishes totally.
May be exactly this circumstances was the reason that some authors stated
the pure one-gluon character of LS-splitting.

4. The first column in tables corresponds to pure perturbation approach. It
is clearly seen that this approach gives only rough qualitative estimate,
but the results are drastically improving with switch on the CIA expansion.
We believe that the using of CIA approach in quark physics will has bright
future. Very convenient for fine-splitting characteristic is the coefficient

\begin{equation}
R=\frac{M\left( ^{3}P_{2}\right) -M\left( ^{3}P_{1}\right) }{M\left(
^{3}P_{1}\right) -M\left( ^{3}P_{0}\right) }.  \label{f20}
\end{equation}

As cited in \cite{1} the experimental values of this parameter are
$R=0.66$ for $b\overline{b}$, $R=0.48$ for $c\overline{c}$ and
also $R=0.21$ for $u \overline{u}$. It is interesting to note that
firstly this parameter was introduced in atomic physics \cite{21}
where its value was established for atoms with two outer electrons
(like Mg) as being $R=2$. Since the fine mass splitting in atoms
is totally due to one-photon exchange this value gives good
indication as to the nature of the Lorentz character of
$q\overline{q}-$ potential. As our calculations show the one-gluon
term is totally vector while non-perturbative confinement term is
of about 75-80\% scalar.

Confinement gives 20-25\% plus to the results, obtained with using only the
one gluon exchange term, for the description of the $LS-$, $ST-$
interaction. Finally we obtain description of fine-effects at the quark
systems with the precision 80-90\%.

These conclusions confirms with the qualitative estimation of
Lucha and Sh\"{o}berl \cite{1}. This result approximately confirms
with the conclusion of Franzini \cite{16} in the part, that the
confinement is pure scalar. We show that the results are very
sensitive to the exact value of $\varepsilon$ and small change of
$\varepsilon$ can destroy the agreement with the experiment what
actually occurred in the icase of Franzini.

We believe that exactly this feature will be successfully
exploited in near future not only in hadronic structure, but also
explaining resonances in $\pi \pi -$ or $\pi N-$ resonances.

\end{document}